\documentclass[12pt,a4paper,russcorr]{article}
\usepackage[dvips]{graphicx}
\usepackage[dvips]{epsfig}

\usepackage[T2A]{fontenc}
\usepackage[cp1251]{inputenc}
\usepackage[english]{babel}

\begin{document}

\begin{center}
{\large\textbf{Transient processes in disordered semiconductor structures\\ under dispersive transport conditions:\vspace{0.2cm}\\ Fractional calculus approach}}

\

\textit{R. T. Sibatov, V. V. Uchaikin}

Ulyanovsk State University

\end{center}

\begin{abstract}
We continue to develop a new approach to description of
charge kinetics in disordered semiconductors. It is based on fractional diffusion equations.
This article is devoted to transient processes in structures under dispersive transport conditions.
We demonstrate that this approach allows us (i) to take into account energetic and topological types of disorder in common, (ii) to consider transport in samples with spatial distributions of localized states, and (iii) to describe transport in non-homogeneous materials with distributed dispersion parameter. Using fractional approach provides some specifications in interpretation of time-of-flight experiments in disordered semiconductors.
\end{abstract}

\


\section{Introduction}

In disordered solids there are often observed significant deviations from classic
transport laws~\cite{Kla:08, Mad:88, Zvy:84, Uch:12}. These phenomena covered by term~\textit{anomalous} or \textit{generalized
transport} are not compatible with the local equilibrium hypothesis.
The non-Gaussian properties of the transport are observed in
amorphous hydrogenated silicon
\cite{Hva:81, Mad:88}, amorphous selenium \cite{Pfi:76, Noo:77},
amorphous chalcogenides \cite{Kolomietz78, Shutov_FTP}, organic
semiconductors, polymers,~\cite{Tut:05, Bassler, Tyutnev},
porous solids \cite{Averkiev00, Averkiev03, Kazakova04},
nanostructured materials \cite{Choudhury03}, polycrystalline
films~\cite{Ramirez_Bon93}, liquid crystals~\cite{Boden}, etc.
Investigation of such processes represents one of the main trends
in contemporary non-equilibrium thermodynamics and statistical
mechanics~\cite{Lebon08}. The results of the generalized transport theory are important for modern applications of disordered materials and mesoscopic systems~\cite{Imr:97, Tess:09, Ruh:11, Shelke:09, Shelke:13}.

Saying about anomalous
transport (AT), we can mean
an unusual value of diffusivity or its time- or space-dependence
in the framework of the standard diffusion approach. When the diffusivity is highly irregular it is more convenient to interpret it as a random field and the process itself  as a complex process consisting of many normal
processes with wide distributions of their characteristics. These processes are denoted by  the term \textit{dispersive transport} (DT).

Numerous experiments manifest the presence of \textit{universal}
DT properties which weakly depend on the detailed atomic and
molecular structure of matter \cite{Sch:75, Jon:96}.
Theoretical foundations of this approach were laid by Scher and Montroll (1975). Their model known as CTRW (Continuous Time Random Walk) has proved to be very fruitful for description of charge kinetics in disordered semiconductors and was followed by a series of articles that used the
waiting time distributions of L\'{e}vy type
\cite{Sch:75}.

From physical point of view, the dispersive transport may be
explained by involving various mechanisms: multiple trapping of
charge carriers into localized states distributed in the mobility gap,
hopping conduction assisted by phonons, percolation
through conducting states, etc. \cite{Sch:75, Ark:83, Zvy:84, Tiedje2, Nik:11}. The variety of approaches reflects
a complexity of the systems and processes under consideration. For this reason,  the construction of a consistent dispersive transport theory based on first principles is still an unsolved problem. Experimental data
revealing universal behavior of some important characteristics of dispersive transport (e.g. time-behaviour of transient photocurrent)
indicates predominance of statistical laws over dynamical ones. Interest in non-Gaussian transport theory has recently revived in connection
with the observation of anomalous relaxation-diffusion processes in nanoscale systems:
nanoporous silicon, glasses doped by quantum dots, quasi-1D systems, and
arrays of colloidal quantum dots. These systems are very promising for applications
in spintronics and quantum computing. They can also be useful for studying the
fundamental concepts of physics of disordered solids: localization, nonlinear effects
associated with long-range Coulomb correlations, occupancy of traps and Coulomb
blockade. Due to the preparation method of colloidal nanocrystals, the energy disorder
is always presented in these systems, which is confirmed by experiments on
fluorescence blinking of single quantum dots (CdSe, CdS, CdSe/ZnS, CdTe, InP,
etc). As shown in some recent papers \cite{Nov:05, Sib:11a}, the
L\'evy statistics plays a crucial role in the interpretation of experiments with charge
transfer in QD arrays.

It is reasonable to believe that kinetic equations describing such
transport processes must have similar forms for different
materials. Nevertheless, the Sher-Montroll version of DT for disordered systems is expressed in form of integral equations while the standard version for ordinary systems has form of partial differential equation. Embedding fractional derivatives in the theory \cite{Nigmatullin2, Wes:91, Metzler1, Metzler2, Mar:00, Barkai, Bisquert1, Bisquert2, Sib:07} removed this unwanted feature and opened opportunities for the development of
normal and anomalous kinetics in the framework of unified
mathematical formalism.

In this paper, we focus on a subclass of DT processes called
\textit{fractional dispersive  transport} (FDT) characterized by involving differential equations of fractional orders [24], which produce  long-tail distributions of the power type. We demonstrate some applications of fractional dispersive transport equations to transient processes in disordered semiconductor structures. In the next section, we list main modifications of the FDT-equation, which describe different situations. Then we briefly consider the time-of-flight methodic, the case of non-uniform distribution of localized states over the sample, and the case of medium with distributed dispersion parameter. We calculate transient process in the diode at dispersive transport conditions. Using fractional approach allows us to provide some specifications in interpretation of the time-of-flight experiment in organic semiconductors. To the end, we consider the influence of topological disorder and percolation on transient current curves.

\section{The family of fractional dispersive transport equations}

Here, we are listing some modifications of  the fractional dispersive transport equation. They contain fractional Caputo and Riemann-Liouville derivatives \cite{Samko, Oldham, UchBook}
$$
_0^\alpha\textsf{D}_t n(\mathbf{r},t)=\frac{1}{\Gamma(1-\alpha)}\int_0^t\frac{\partial
n(r,t')/\partial t'}{(t-t')^\alpha} dt', \qquad
_{0}\textsf{D}_t^{\alpha}=\frac{1}{\Gamma(1-\alpha)}\frac{\partial}{\partial
t}\int_0^t\frac{n(r,t')}{(t-t')^\alpha} dt'.
$$

$\bullet$ The fractional Fokker-Planck equation\footnote{As known (see Ref.~\cite{Met:99}) the fractional Fokker-Planck equation is consistent
with the fluctuation-dissipation theorem and the generalized Einstein relation. In the absence of external fields, the dispersive diffusion packet spreads slower than in normal case: $\Delta\propto t^{\alpha/2}$ ($\alpha<1$). The fractional time derivative caused by wide distributions of waiting times with power tails leads to additional dispersion of the particle coordinates.} for the total concentration of nonequilibrium carriers
$n(\textbf{r},t)$ (see details in \cite{Uch:12}):
\begin{equation}\label{eq_total}
_0^\alpha\textsf{D}_t n(\mathbf{r},t)+
\mathrm{div}\left(\frac{}{} \mathbf{K}\ n(\mathbf{r},t)-C\nabla
n(\mathbf{r},t)\right)=\frac{t^{-\alpha}}{\Gamma(1-\alpha)}n(\mathbf{r},0)
\end{equation}
can be used for hopping conduction over the Poisson ensemble of traps and for multiple trapping into band tail states with the  exponential energy distribution \cite{Barkai, Sib:07, Uch:08}. Here, $0<\alpha\leq1$ is the dispersion parameter, $\mathbf{K}\propto \mathbf{E}$ the anomalous advection coefficient, and $C$ the anomalous diffusion coefficient. For the multiple trapping mechanism, the absolute value of $\mathbf{K}$ is expressed through microscopic parameters as $K=c^\alpha l$, where ${c={w}_0[\sin(\pi\alpha)/\pi\alpha]^{1/\alpha}}$, ${w}_0$
is the capture rate of carriers into localized states,
$\mu$ and $D$ are mobility and diffusion coefficient of delocalized carriers, $\mathbf{K}=\tau_0 c^\alpha \mu\mathbf{E}$ is a dispersive advection, $C=\tau_0 c^\alpha D$ is an anomalous diffusion coefficient. Parameters $\tau_0$ and $l$ are the average  time and length of delocalization, respectively.
For variable range hopping,
${c=\nu_0[\sin(\pi\alpha)/\pi\alpha]^{1/\alpha}}$,
where $\nu_0$ is the characteristic rate of jumps between the traps.

Solutions $n_\alpha(\mathbf{r},t)$ of fractional equation (\ref{eq_total}) are expressed through the solutions
$n_1(\mathbf{r},t)$ of the ordinary Fokker-Planck equation by the relation \cite{Uch:99, Barkai}:
\begin{equation}\label{sol_eq_time1}
n(\mathbf{r},t)=\int_0^t d\tau\ \left\{\frac{c t}{\alpha
\tau}{\left(\frac{\tau}{\tau_0}\right)}^{-1/\alpha}g_+\left(c
t{\left(\frac{\tau}{\tau_0}\right)}^{-1/\alpha};\alpha\right)n_1(\mathbf{r},\tau)\right\},
\end{equation}
where $g_+(t;\alpha)$ is the one-sided L\'evy stable pdf~\cite{Stable},
which can be determined by its Laplace transform
$$
\int\limits_0^\infty e^{-\lambda t} g_+(t;\alpha)\ dt=e^{-\lambda^\alpha}.
$$

 Eq.~(\ref{sol_eq_time1}) allows to find analytical solutions in simple cases and to derive the general Monte Carlo algorithm \cite{Sib:09}.

$\bullet$ The equation for the density of delocalized carriers $n_d(\textbf{r},t)$  in case of multiple trapping has the form
\begin{equation}\label{eq_FDT2}
\frac{\partial n_d(\mathbf{r},t)}{\partial t}+\frac{l}{\tau_0 K}\
_0\textsf{D}_t^{\alpha}\
n_d(\mathbf{r},t)+\mathrm{div}\left(\frac{}{}\mu\mathbf{E}\
n_d(\mathbf{r},t)-D\nabla n_d(\mathbf{r},t)\right)=0.
\end{equation}

$\bullet$ The transport equation taking into account the recombination of localized carriers~is derived~\cite{UchSib:12_WS, Uch:12, Sib:09} in the form:
$$
e^{-\gamma t}\
_0^\alpha\textsf{D}_t\
e^{\gamma t}\ n(\mathbf{r},t)
+(c^\alpha\tau_0)\mathrm{div}\left[\frac{}{} \mu\mathbf{E}\ n-D
\nabla n\right]=0,
$$
where $\gamma$ is a recombination rate for localized carriers.

$\bullet$ The fractional dispersive transport equation taking into account the monomolecular recombination of delocalized carriers is obtained in Ref.~\cite{Sib:09}:
\begin{equation}\label{eq_FDT_monomol}
\frac{\partial n(\mathbf{r},t)}{\partial t}+\frac{\tau_{0} K}{l} \
_{0}\textsf{D}_t^{1-\alpha} \left[\ \mathrm{div}\left(
-\mu\mathbf{E}\ \delta n(\mathbf{r},t)-D\nabla
n(\mathbf{r},t)\frac{}{}\right)+\frac{\delta
n(\mathbf{r},t)}{\tau_\mathrm{mr}}\right]=0,
\end{equation}
where $\tau_\mathrm{mr}$ is the monomolecular recombination time, and $\delta n$ the concentration of non-equilibrium carriers.

$\bullet$ The fractional formalism has allowed to derive the bipolar diffusion equation for dispersive transport in case of multiple trapping \cite{Sib:09}
\begin{equation}\label{eq_FDT_bipolar}
\left[\frac{\sigma_n}{\sigma}\ _0\textsf{D}^{\alpha_p}_t +
\frac{\sigma_p}{\sigma}\ _0\textsf{D}^{\alpha_n}_t\right]p_d +
\mu_{\mathrm{amb}}\ \mathbf{E}\ \nabla p_d \ - D_{\mathrm{amb}}
\nabla^2 p_d+\frac{\delta p_d}{\tau^*}= 0.
\end{equation}
Here, $\sigma_n=\mu_n n_d,\ \sigma_p=\mu_p p_d$ are conductivities of delocalized electrons and holes, $\sigma=\sigma_n+\sigma_p$;
$\mu_{\mathrm{amb}}=\mu_p^* \mu_n^*(n_d-p_d)[\mu_n^* n_d+\mu_p^*
p_d]^{-1} $ is bipolar dispersive drift mobility, and
$D_{\mathrm{amb}}=(\mu_n^* n D_p^*+\mu_p^* p D_n)(\mu_n^*
n+\mu_p^* p)^{-1}$ bipolar diffusion coefficient. The fractional bipolar transport equation contains two
fractional derivatives of different orders in the general case. This is a particular case of distributed order equation.

$\bullet$ In case of distributed dispersion parameter, the transport equation is of the form~\cite{Sib:09}
\begin{equation}\label{eq_FDT_distributed}
 \frac{\partial n(\mathbf{r},t)}{\partial
t}+\ \int_0^1 d\alpha \ \rho(\alpha)\ _{0}\textsf{D}_t^{1-\alpha}
\mathrm{div}\left( -\mathbf{K}_\alpha\ \delta n-C_\alpha\nabla
n\right)=0.
\end{equation}
Here, $\rho(\alpha)$ is the distribution density of dispersion parameter.

$\bullet$ For exponentially truncated power law distributions of localization times in the generalized Scher-Montroll model,
\begin{equation}\label{eq_FDT_trunc}
\frac{\partial n(\mathbf{r},t)}{\partial t}
+\mathrm{div}\left[e^{-\gamma t}\ {_0\textsf{D}_t^{1-\alpha}}
e^{\gamma t}\ \left( \mathbf{K}\  n(\mathbf{r},t)- C\ \nabla
n(\mathbf{r},t)\right)\right]=0,
\end{equation}
where $\gamma$ is a truncation parameter.
In this case, localization (waiting) times have a finite variance, the Central Limit Theorem is applicable in this case, transport at large times is normal. The transition from the dispersive regime to the Gaussian one in the time-of-flight experiment is theoretically described on the base of the truncated L\'evy statistics in Ref.~\cite{Sib:11}.

$\bullet$ In frames of the multiple trapping model, the equation for the delocalized carrier concentration in case of arbitrary density of states $\rho(\varepsilon)$ and percolative nature of conduction ways, is obtained in the form~\cite{Uch:12}:
$$
\frac{\partial n_d(\mathbf{r},t)}{\partial
t}+\tau_0^{-1} \tau_\beta^\beta\ _0\textsf{D}_t^\beta n_d(\mathbf{r},t)+\tau_0^{-1}\frac{\partial}{\partial t}\int_0^t dt'\
n_d(\mathbf{r},t-t')\
\int_0^{\varepsilon_g}d\varepsilon\ \rho(\varepsilon)
\exp\left(-{w}_\varepsilon t'\
\mathrm{e}^{-\varepsilon/kT}\right) +
$$
\begin{equation}\label{eq_multiple_gen_case}
+ \mathrm{div}[\ \mu\mathbf{E}\
n_d(\mathbf{r},t)-D\nabla n_d(\mathbf{r},t)]=0.
\end{equation}
The term with fractional derivative of order $\beta$ is consistent with the comb model of a percolation cluster \cite{Arkh:91}. The constant $\tau_\beta$ is the characteristic residence time in "dead bonds" of a percolation cluster.
For hopping in a medium with the Gaussian energetic density of states, the equation for the carrier concentration $n_\mathrm{eff}(x,t)$ near the transport layer is as follows \cite{Uch:12}:
$$
\frac{a_\beta}{c^\beta}\ _0\textsf{D}_t^\beta{n_\mathrm{eff}}(x,t)+a_\alpha A\frac{\partial}{\partial t}\int\limits_0^t \frac{{n_\mathrm{eff}}(x,t')}{(t-t')^{kT/\sigma}} \exp {\left\{-\left(\frac{\sigma (t-t')}{4kT}\right)^{kT/2\sigma}\right\}}dt'=
$$
\begin{equation}\label{eq_FDT_gauss}
=  l \  \frac{\partial {n_\mathrm{eff}}(x,t)}{\partial x}- D^*\ \frac{\partial^2 {n_\mathrm{eff}}(x,t)}{\partial x^2}=0.
\end{equation}
The second term describes thermally activated hops between localized states distributed with Gaussian density, i.e. $\propto \exp(-\varepsilon^2/2\sigma^2)$.

\section{Photocurrent decay in the time-of-flight experiment}

In classical ``time-of-flight'' experiments, electrons and
holes are usually generated in a sample by a pulse of laser
radiation from the side of the semitransparent electrode. The
voltage applied to the electrodes is such that the corresponding
electric field inside the sample is significantly stronger than
the field of nonequilibrium charge carriers. The electrons
(or holes, depending on the voltage sign) enter the
semitransparent electrode, while holes (or electrons) drift to the
opposite electrode. In the case of normal transport,
drifting carriers in the field $E$ give rise to a rectangular
photocurrent pulse:
\begin{equation}\label{experimental_time_T2}
I(t)\propto  \left\{
\begin{array}{cc}
{\rm const},& t<t_{T},\\
0,& t>t_{T},\\
\end{array} \quad \alpha<1,
\right.
\end{equation}
where the time of flight $t_{T}$ is given by drift velocity
$v_d$ and sample length $L$: $t_{T}=L/v_d$. Taken together, the
scattering of delocalized carriers during the drift, trapping into
localized states, and thermal emission of the carriers lead to
packet spreading. Such a packet has a Gaussian shape with a mean
value of $\langle x(t)\rangle\propto t$ and width $\Delta
x(t)\propto \sqrt{t}$. In this case, the transient current $I(t)$
remains constant until the leading edge of the Gaussian packet
reaches the opposite edge of the sample. The current decrease takes a
time of $ \Delta x/\langle v_d\rangle$. As a result, the right edge
of the photocurrent pulse becomes smooth. Such
a picture is typical for most ordered materials.

However, when determining drift mobility in certain disordered
(amorphous, porous, disordered organic, strongly doped, etc.)
semiconductors, a specific signal of transient current $I(t)$,\index{transient current} is
observed, having two regions with the power-law behavior of $I(t)$
and an intermediate region:
\begin{equation}\label{anomal}
I(t) \propto  \left\{
\begin{array}{cc}
t^{-1+\alpha},& t<t_{T},\\
t^{-1-\alpha},& t>t_{T},\\
\end{array} \quad \alpha<1.
\right.
\end{equation}

Exponent $\alpha$, termed the \emph{dispersion parameter}\index{dispersion parameter}, depends on the
medium characteristics and can vary with temperature. Parameter
$t_{T}$ is called \emph{transient time}\index{transient time} (or \emph{time of flight})\index{time of flight} in analogy with
normal transient processes, but has a different physical sense. It
has been shown experimentally \cite{Sch:75, Pfi:76} that in the dispersive transport regime the
following relationship takes place:
\begin{equation}\label{experimental_time_T}
t_{T}\propto (L/U)^{1/\alpha},
\end{equation}
where $U$ is the voltage.

As noted in Refs.~\cite{Sch:75, Pfister} the shape of the transient
current signal in the reduced coordinates $\lg[I(t)/I(t_{T})]$
-- $\lg[t/t_{T}]$ is virtually independent of the applied
voltage and sample size. This property, inherent in many (but not
all: see \cite{Roosebroeck}), materials, is referred to as the
property of shape universality of transient current curves. Occurrence of these features in many
disordered materials confirms the universality of transport
properties. A large number of experimental observations of this
universality were reported both in early and recent publications
(see for details Refs.~\cite{Mad:88, Zvy:84, Sch:75, Jon:96, Tut:05}).

The transient photocurrent $I(t)$ in a sample of the length $L$ is determined through the conductivity
current density as
\begin{equation}\label{eq_trcurrent_gen}
I(t)=(1/L)\int_0^L j(x,t)dx.
\end{equation}
and related to the one-dimensional concentration of injected carriers $n(x,t)$ by the following relation:
\begin{equation}\label{eq_transient_current}
I(x,t)=\frac{e}{L}\frac{d}{dt}\int\limits_0^L (x-L)\ n(x,t) dx.
\end{equation}

Rewriting equation (\ref{eq_total2}) in the one-dimensional form and neglecting by the diffusion component, we arrive at the equation
\begin{equation}\label{eq_total2}
_0^\alpha\textsf{D}_t n(x,t)+
K \frac{\partial}{\partial x} n(x,t)=N\delta(x)\frac{t^{-\alpha}}{\Gamma(1-\alpha)},
\end{equation}
which has the following solution
\begin{equation}\label{total_concentration}
n(x,t)=\frac{N t}{\alpha
K}{\left(\frac{x}{K}\right)}^{-1/\alpha-1} g_+\left(t
{\left(\frac{x}{K}\right)}^{-1/\alpha};\alpha\right).
\end{equation}
Here, $N$ is a surface density of injected carriers. Substituting the latter function into Eq.~(\ref{eq_transient_current}), we arrive at the expression for the transient current density:
\begin{equation}\label{photocurrent2}
I(t)=\frac{e K N \alpha}{L
}{t}^{\alpha-1}\int_{\zeta_0}^{\infty}\zeta^{-\alpha}g_+(\zeta;\alpha)\
d\zeta,\quad \zeta_0=t\ {(L/K)}^{-1/\alpha}.
\end{equation}

\begin{figure}[tbhp]
\centering
\includegraphics[width=0.5\textwidth]{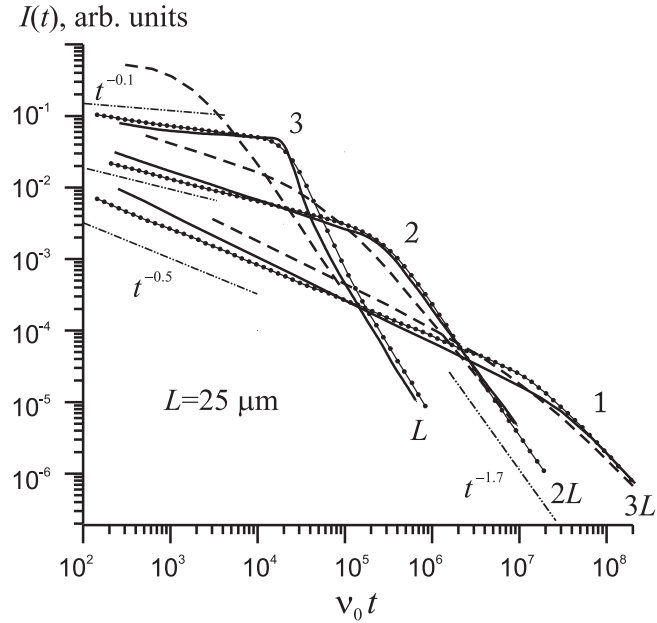}
\caption{Transient current curves. Dashed lines are the solutions obtained within the Arkhipov-Rudenko $\tau$-approximation, solid lines are the solutions of the Arkhipov-Rudenko-Nikitenko equation (digitized from Ref.~\cite{Nik:07}), dotted lines are calculated through the solutions~(\ref{sol_eq_time1}) of Eq.~(\ref{eq_total}). The parameters are specified in the text.}\label{fig_nikitenko1}
\end{figure}

 The transient current curves calculated by Eq.~(\ref{photocurrent2}) are presented in Fig.~\ref{fig_nikitenko1} in comparison with solutions of the Arkhipov-Rudenko $\tau$-approximation \cite{Arkh:82} and the Arkhipov-Rudenko-Nikitenko diffusion equation \cite{Nik:07}. The following parameters have been taken for calculations: $E=5\cdot10^5$~V/cm, ${w}_0=10^6$~c$^{-1}$, $\mu_0\tau_0=2.5\cdot10^{-16}$~m$^2/$V, $l=12,5$~nm. Parameters of fractional equations: 1)~$\alpha=0.5$, $K=8$~$\mu$m/s$^{0.5}$, $L=75$~$\mu$m; 2)~$\alpha=0.7$, $K=73$~$\mu$m/s$^{0.7}$, $L=50$~$\mu$m; 3)~$\alpha=0.9$, $K=343$~$\mu$m/s$^{0.9}$, $L=25$~$\mu$m.

 In case of truncated waiting time distributions, equation (\ref{eq_FDT_trunc}) leads to the following expression for the conduction current
  \begin{equation}\label{current_density_trunc}
j(x,t)= e N
\exp\left[\frac{x}{K}\gamma^\alpha-\gamma
t \right] \ \left(\frac{x}{K}\right)^{-1/\alpha}
g_+\left( t \left(\frac{x}{K}\right)^{-1/\alpha};\alpha\right),
\end{equation}
where $\gamma$ is the truncation parameter. Transformation of transient current
curves with an increasing~$L/l$ ratio is studied in \cite{Sib:11}. If the transient time $t_T$ is much smaller than the truncation time~$\gamma^{-1}$ the
transport remains dispersive and does not pass to Gaussian
asymptotics. For $t_{T}\gg \gamma^{-1}$, transport in the long-time asymptotic regime becomes normal.

\begin{figure}[htb]
\centering
\includegraphics[width=0.65\textwidth]{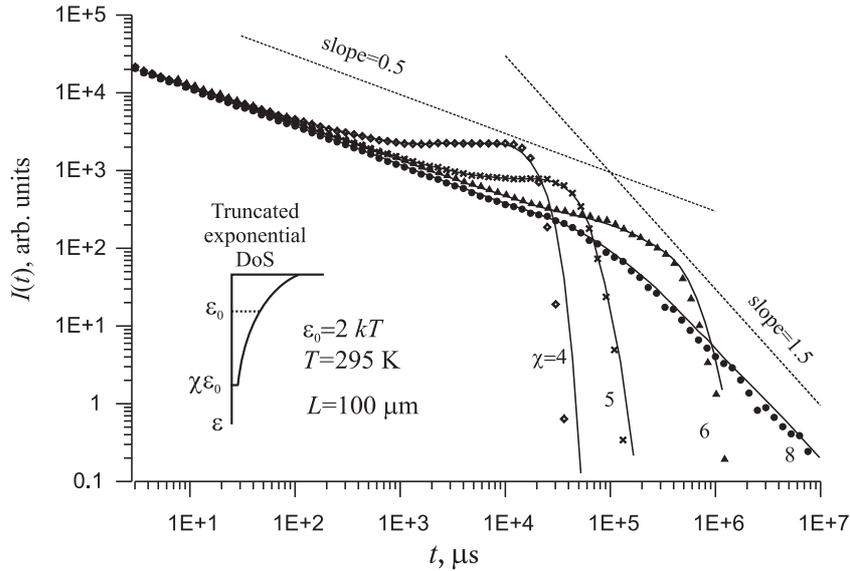}
\caption{Transient current curves in case of the bounded exponential spectrum of localized states. Points are the results of Monte Carlo simulation of multiple trapping, lines represents calculations according to Eqs.~(\ref{eq_trcurrent_gen}) and (\ref{current_density_trunc}).}\label{fig_tr_exp_trunc}\vspace{0.5cm}
\end{figure}

Fig.~\ref{fig_tr_exp_trunc} presents transient current curves, calculated by inserting function (\ref{current_density_trunc}) into Eq.~(\ref{eq_trcurrent_gen}). This behavior can be caused by truncation of exponential density of localized states in the multiple trapping model. The comparison is presented in Fig.~\ref{fig_tr_exp_trunc}. Parameters of multiple trapping model: $\varepsilon_0=2 kT$, $T=295$~K, $L$=100$~\mu$m.
In case of non-truncated exponential density of localized states, dispersive transport ($\alpha=kT/\varepsilon_0=0.5$) is observed. The truncation of $\rho(\varepsilon)$ at $\chi\varepsilon_0$ transforms transient current curves not only in the tail $t\gg t_T$, it can lead to the appearance of a plateau. In this case, the power law tail of current is not observed. We can meet another situation when the plateau and the power law tail are presented together in curves. This fact can not be explained by the boundedness of the band tail. Possible explanations are given below.

\subsection{Non-uniform spatial distribution of localized states}

Consider the case of inhomogeneous spatial distribution of localized states. When traps are distributed over a sample with the density $\rho(x)$, the average number of localization events for one carrier in a layer of thickness $x$ is equal to $k=\int_0^x \rho(x) dx$,
and the conduction current density has the form
\begin{equation}\label{nonhomo_conduction_current}
j(x,t)=e N \left[\int_0^x \rho(x) dx\right]^{-1/\alpha}\
g_+\left(c t \left[\int_0^x \rho(x)
dx\right]^{-1/\alpha};\alpha\right).
\end{equation}
Here $c$ is the scale parameter of  the localization time distribution: $\textsf{Prob}(T>t)\sim (c t)^{-\alpha}/\Gamma(1-\alpha)$, $t\rightarrow\infty$.
From the continuity equation
$$
\frac{\partial n(x,t)}{\partial t}+\frac{1}{e}\frac{\partial
j(x,t)}{\partial x}=\delta(x)\delta(t),
$$
one can find the total concentration of carriers,
\begin{equation}\label{nonhomo_f_x_t}
n(x,t)=-\frac{1}{e}\frac{\partial}{\partial x}\int\limits_0^t
j(x,t) dt
=ct\ \alpha^{-1} \left[\int\limits_0^x \rho(x)
dx\right]^{-1/\alpha-1}\ \rho(x)\ g_+\left(c t
\left[\int\limits_0^x \rho(x) dx\right]^{-1/\alpha};\alpha\right),
\end{equation}
and the transient current,
$$
I(t)=\frac{e N}{L}\int_0^L dx \ \left[\int_0^x \rho(x) dx\right]^{-1/\alpha}\
g_+\left(c t \left[\int_0^x \rho(x)
dx\right]^{-1/\alpha};\alpha\right).
$$

\begin{figure}[tbh]
 \centering
 \includegraphics[width=1\textwidth]{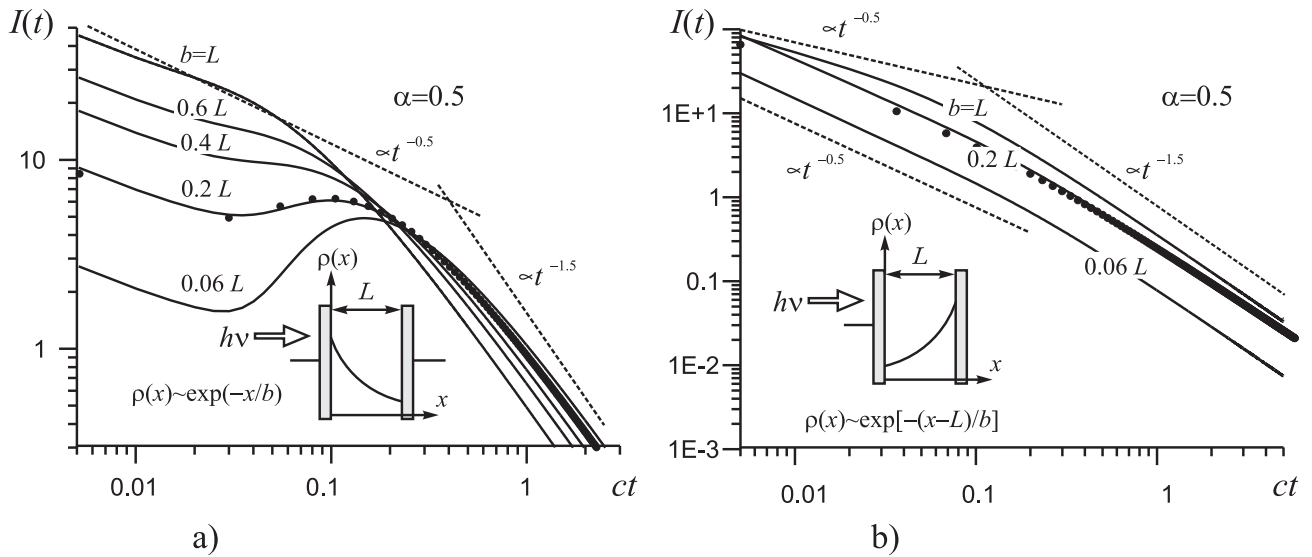}
 \caption{a) Dispersive transient current curves ($\alpha=0.5$) in case of non-uniform spatial distribution of localized states
  $\rho(x)\propto\exp(-x/b)$ for different $b$ values. b) The same for $\rho(x)\propto\exp[-(x-L)/b]$. Points are the results of numerical simulation for $b=0.2 L$.}\label{fig_surf_layer}
 \end{figure}

Different types of spatial distribution of localized states are considered in Refs.~\cite{Ryb:89, Uch:12, Sib:12}. Fractional approach confirms results obtained in \cite{Ryb:89}.
Take a look at Fig.~\ref{fig_surf_layer} showing the transient current curves in the case of surface layers depleted or enriched by traps, exponential distributions of traps over the sample have been taken. In the first case we observe the appearance of a maximum on the curves, in the second case we obtain a more diffuse characteristics than in the case of homogeneous distribution of traps in the sample. Analytical results are in accordance with the Monte Carlo simulation of the transport by multiple trapping.

\begin{figure}[hbp]
 \centering
 \includegraphics[width=1\textwidth]{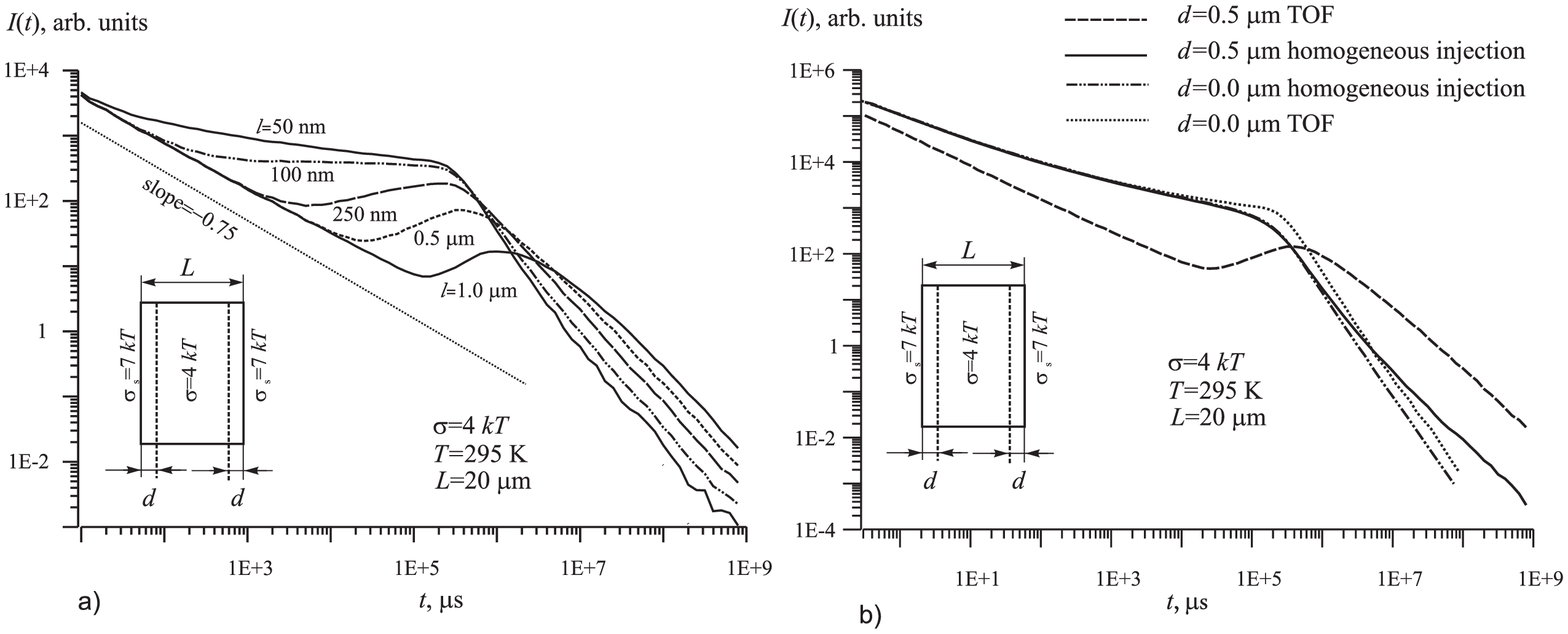}
 \caption{Transient currents with taking into account surface layers of different thickness ($T=295$~K, $L=20$~mkm,  $E=10^6$~V/cm, $\hat{\sigma}=4$ is the parameter of Gaussian disorder of the bulk, and $\hat{\sigma}=7$ of the surface layers. b) Influence of surface layers on transient current curves in the time-of-flight method (near electrode generation of carriers), and in the case of uniform generation of carriers.}\label{fig_threelayer}
 \end{figure}

The influence of surface layers can be analyzed by considering the three-layer structure \cite{Tut:05}. The outer layers are surface layers and the main bulk of the material is located between them. Here, barrier effects are neglected, that is correct for large voltages applied to the structure. Calculation has been performed for the case of hopping in a material with Gaussian energetic disorder ($\hat{\sigma}=\sigma/kT$). Transient current in each layer can be found from Eqs.~(\ref{eq_FDT_gauss}, \ref{eq_transient_current}), the total current is calculated as $I_1(t)+I_2(t)+I_3(t)$. In Fig.~\ref{fig_threelayer}, transient current curves generated by surface (time-of-flight method) and uniform injection of carriers into the three-layer system are presented. These calculations show that appearance of a hill on transient current curves can be explained by the presence of disordered surface layers. This result is consistent with the calculations in frames of the Arkhipov-Rudenko $\tau$-formalism~\cite{Tut:05}.

\subsection{B\"assler's model of Gaussian disorder}

The Scher-Montroll  approach \cite{Sch:75} and the Arkhipov-Rudenko theory \cite{Arkh:82} predict a transition to the Gaussian regime, when the dispersion parameter $\alpha$ tends to 1.
In the framework of multiple trapping and thermoactivated hops, transition to the normal statistics is observed when temperature is increased. However, it should be noted that the model for hopping transport in organic semiconductors predicts the transition to the normal transport, when sample thickness is increased or applied voltage is decreased \cite{Bassler}.
In other words, a change in transport statistics can be due to changes in macroscopic large-scale parameters. For small transient times, i.e. small values of sample thickness and/or high voltages, the normalized transient current curves are almost universal, and correspond to the dispersive mode of transport. In samples with greater thickness, or at lower voltages, a plateau on the curves of $I(t)$  is observed~\cite{Bassler, Tut:05}, which indicates the Gaussian mode of transfer. This phenomenon demonstrating the spatio-temporal scale effect relates to the case with many low molecular weight, molecular-doped and conjugated polymers and can be described in terms of the theory of quasi-equilibrium transport~\cite{Nik:11}.

B\"assler's model assumes that the energy distribution of hopping
centers involved in tunnel-activation transfer is described by the Gaussian function. In this case, waiting time
distribution have truncated power law form. All moments of sojourn
times are finite, and normal transport regime has to be observed
at large times.

Detailed analysis of the localization time distributions \cite{Uch:12} shows that the complementary cumulative function $\overline{\Psi}(t)={\rm Prob}(\tau> t )$ in the B\"assler model can be described by an inverse power function multiplied by a stretched exponential one. It is also important that the index of the stretched exponential function is not arbitrary: it is twice smaller than the power law index $ \alpha_1 = kT/\sigma $:
$$
\overline{\Psi}(t)=\left\langle\exp\left(-\nu_0 t\
\mathrm{e}^{-\varepsilon/kT}\right)\right\rangle=\int\limits_{-\infty}^{+\infty} g(\varepsilon)
\exp\left(-\nu_0 t\
\mathrm{e}^{-\varepsilon/kT}\right)d\varepsilon
$$
$$
\approx \overline{\Psi}^{\mathrm{as}}(t)\equiv At^{-kT/\sigma}\exp\left(-(\sigma t/4 kT)^{kT/2\sigma} \right).
$$

It is worth to note that waiting time distributions and transient current curves obtained in frames of the multiple trapping model are in agreement with the results of direct simulation for the hopping mechanism \cite{Har:96}. This means that in not too strong electric fields, the macroscopic manifestations of both mechanisms are indistinguishable, despite their significant physical difference. In the opinion of Hartenstein et al. \cite{Har:96}, the cause of this lies in the existence of the transport energy level in the hopping model. The transport level plays a role of the mobility edge \cite{Gru:79, Sha:85, Mon:85, Nik:11}.

\subsection{Distributed dispersion parameter}

Transient current relaxation in certain disordered semiconductors, for example, porous silicon \cite{Bisi}, assumes the form
\begin{equation}\label{transient_current_porous}
I(t) \propto  \left\{\begin{array}{cc}
t^{-1+\alpha_i},& t<t_{T},\\
t^{-1-\alpha_f},& t>t_{T},\\
\end{array} \quad 0<\alpha_i\neq\alpha_f<1.
\right.
\end{equation}
The Scher-Montroll model of charge transport in disordered
semiconductors leads to the current dependence (\ref{anomal}), where
$\alpha_i = \alpha_f = \alpha$. As shown in Ref.~\cite{Averkiev},
the value of $\alpha$ found from the dependence of carrier flight time in porous
silicon on the electric field strength does not coincide with that
determined from transient photocurrent curves. The authors explain this fact assuming additional
dispersion in terms of carrier mobility in structurally
inhomogeneous porous silicon samples. It seems quite natural to
extend this idea by involving dispersion of the
parameter $\alpha$. As will be seen below, this assumption is
enough to substantiate dependence~(\ref{transient_current_porous}), at least for the
discrete spectrum $\{\alpha_1,\ \alpha_2,\ \dots,\ \alpha_m\}$.

Let $k_j$ be a portion of traps that capture carriers for random
time $\tau$ distributed according to an asymptotically power law
with exponent $\alpha_j$. The distribution of waiting times averaged over $\alpha$ has the form
$$
\Psi(t)\sim 1-\sum_j k_j \frac{{(b_j\
t)}^{-\alpha_j}}{\Gamma(1-\alpha_j)},
$$
where $b_j$ are normalization constants. The relationship between concentrations of localized and quasi-free carriers takes now the form
\begin{equation}\label{relation4}
\frac{\partial n_t(\mathbf{r},t)}{\partial
t}=\frac{1}{\tau_0}\sum_j c_j^{-\alpha_j}\
_0\textsf{D}_t^{\alpha_j} n_d(\mathbf{r},t),\qquad c_j=b_j\ {(k_j)}^{-1/\alpha_j}.
\end{equation}
Combined with the continuity equation, expression (\ref{relation4}) gives the drift-diffusion equation for the concentration of
delocalized carriers in the case of the discretely distributed
dispersion parameter:
$$
\frac{\partial n_d(\mathbf{r},t)}{\partial
t}+\frac{1}{\tau_0}\sum_j c_j^{-\alpha_j}\
_0\textsf{D}_t^{\alpha_j}  n_d(\mathbf{r},t')+
$$
\begin{equation}\label{eq_distributed_alpha}
+\mathrm{div}\left(\frac{}{}\mu\mathbf{E}\
n_d(\mathbf{r},t)-D\nabla
n_d(\mathbf{r},t)\right)=n(\mathbf{r},0)\ \delta(t).
\end{equation}

To calculate the transient current governed by the latter equation, we neglect diffusion, regard the electric field as
being uniform, and align the $x$-axis along field $\mathbf{E}$. Then,
equation (\ref{eq_distributed_alpha}) can be rewritten as
$$
\frac{\partial n_d(x,t)}{\partial
t}+\frac{1}{\tau_0}\sum_j c_j^{-\alpha_j}\
_0\textsf{D}_t^{\alpha_j} n_d(x,t)+\mu E\ \frac{\partial n_d(x,t)}{\partial
x}=N\delta(x)\ \delta(t).
$$
The Laplace transform
$$
\widetilde{n}_d(x,s)=\int_0^\infty dt\
n_d(x,t) \exp(-s t)
$$
satisfies the equation
$$
s \widetilde{n}_d(x,s)+\frac{1}{\tau_0}\sum_j
{\left(\frac{s}{c_j}\right)}^{\alpha_j}\widetilde{n}_d(x,s)+\mu
E\ \frac{\partial \widetilde{n}_d(x,s)}{\partial
x}=N\delta(x),
$$
solution of which (for the case $\alpha_j<1$) has the form
$$
\widetilde{n}_d(x,s)=\frac{N}{\mu E A}\
\exp\left[-\frac{x}{\mu E \tau_0}\sum_j
{\left(\frac{s}{c_j}\right)}^{\alpha_j}\right],\qquad \mu E
\tau_0=l,
$$
with $A$ standing for  the sample area transverse to the electric field.

 \begin{figure}[htb]
 \centering
 \includegraphics[width=0.7\textwidth]{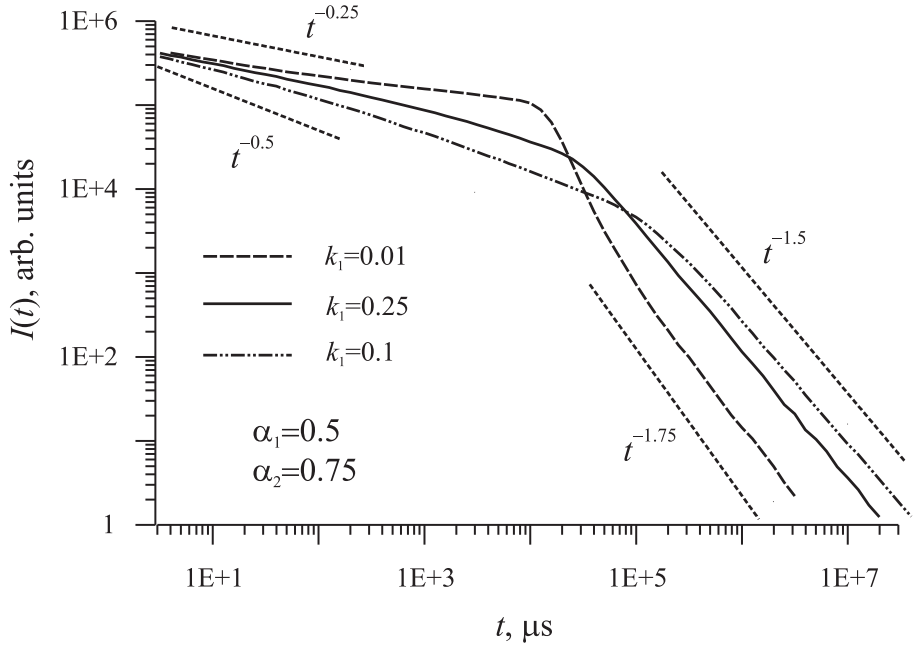}
 \caption{Transient current curves for dispersive transport characterized by two dispersion parameters $\alpha_1=0.5$ and $\alpha_2=0.75$. $\mu E \tau_0=10$~nm, $E=10^6$~V/cm. Fractions $k_1$ of first kind traps ($\alpha_1=0.5$) are indicated in figure, $k_2=1-k_1$. Dashed lines corresponds to
 power laws with exponents determined by dispersion parameters, $t^{-1\pm\alpha}$.}\label{fig_distr_order}
 \end{figure}

On assumption that traps are uniformly distributed over the
sample, the time transform of the total charge carrier density $n(x,t)$
for $x\gg l$ is written as
\begin{equation}\label{f_x_lambda}
\widetilde{n}(x,s)=\frac{N}{l
s}\sum_j(s/c_j)^{\alpha_j}\exp\left(-\frac{x}{l} {\sum}_{j}
{(s/c_j)}^{\alpha_j}\right).
\end{equation}
For the Laplace image of the transient current, we have:
\begin{equation}\label{tr_cur_transformant}
\widetilde{I}(s)=\frac{e N l}{L}\ \frac{1-\exp\left(-L\sum_j
{(s/c_j)}^{\alpha_j}/l\right)}{\sum_j{(s/c_j)}^{\alpha_j}}.
\end{equation}

In order to see the long-time dependence of transient current, one should apply the Tauberian
theorem, according to which the behavior of function $I(t)$ for
$t\gg c_j^{-1}$ is
determined by that of function
(\ref{tr_cur_transformant}) for $s\ll c_j$:
$$
\widetilde{I}(s)\sim \frac{e N l}{L}\ \frac{2 L\sum_j
{(s/c_j)}^{\alpha_j}/l-{\left(-L\sum_j
{(s/c_j)}^{\alpha_j}/l\right)}^2}{2\sum_j{(s/c_j)}^{\alpha_j}}\sim
eN-\frac{e N L}{2 l}{(s/b_{\mathrm{min}})}^{\alpha_{\mathrm{min}}}.
$$
Here $\alpha_{\mathrm{min}}$ is the minimum value from the set
$\{\alpha_1, \alpha_2, \dots, \alpha_m\}$ and $b_\mathrm{min}$ is the corresponding value of the normalization
constant. The inverse Laplace transformation leads to
$$
I(t)\propto t^{-1-\alpha_\mathrm{min}},\quad{t\gg c_j^{-1}}.
$$
In the case of $s/c_j\gg (l/L)^{1/\alpha_j}$ for all $j$, it follows that
$$
\widetilde{I}(s)\sim \frac{e N l}{L
\sum_j{(s/c_j)}^{\alpha_j}}\sim\frac{e N l}{L
{(s/b_\mathrm{max})}^{\alpha_\mathrm{max}}}, \quad s\gg c_j,
$$
where $\alpha_\mathrm{max}$ is the maximum value from the set
$\{\alpha_1, \alpha_2, \dots, \alpha_m\}$ and $b_\mathrm{max}$ is the corresponding value of the
normalization constant. Hence follows
$$
I(t)\propto t^{-1+\alpha_\mathrm{max}}, \quad t\ll c_j^{-1}.
$$

Thus, if the exponent in the carrier residence time
distribution in traps takes on one of the values from an
ordered set $\{\alpha_1, \alpha_2, \dots,
\alpha_m\}$ (discrete spectrum), the transient
current behavior is determined by the maximum value of
$\alpha_\mathrm{max}=\alpha_m$ in the initial time segment, and by the minimum
value of $\alpha_\mathrm{min}=\alpha_1\neq\alpha_m$ (Fig.~\ref{fig_distr_order}) in the terminal one, in agreement with
the results of the aforementioned experiments.

In Fig.~\ref{fig_distr_order}, there are transient current curves for dispersive transport characterized by two dispersion parameters $\alpha_1=0.5$ and $\alpha_2=0.75$. $\mu E \tau_0=10$~nm, $E=10^6$~V/cm. Fraction of the first type traps is $k_1$, $k_2=1-k_1$.

\begin{figure}[hbt]
\centering
\includegraphics[width=1\textwidth]{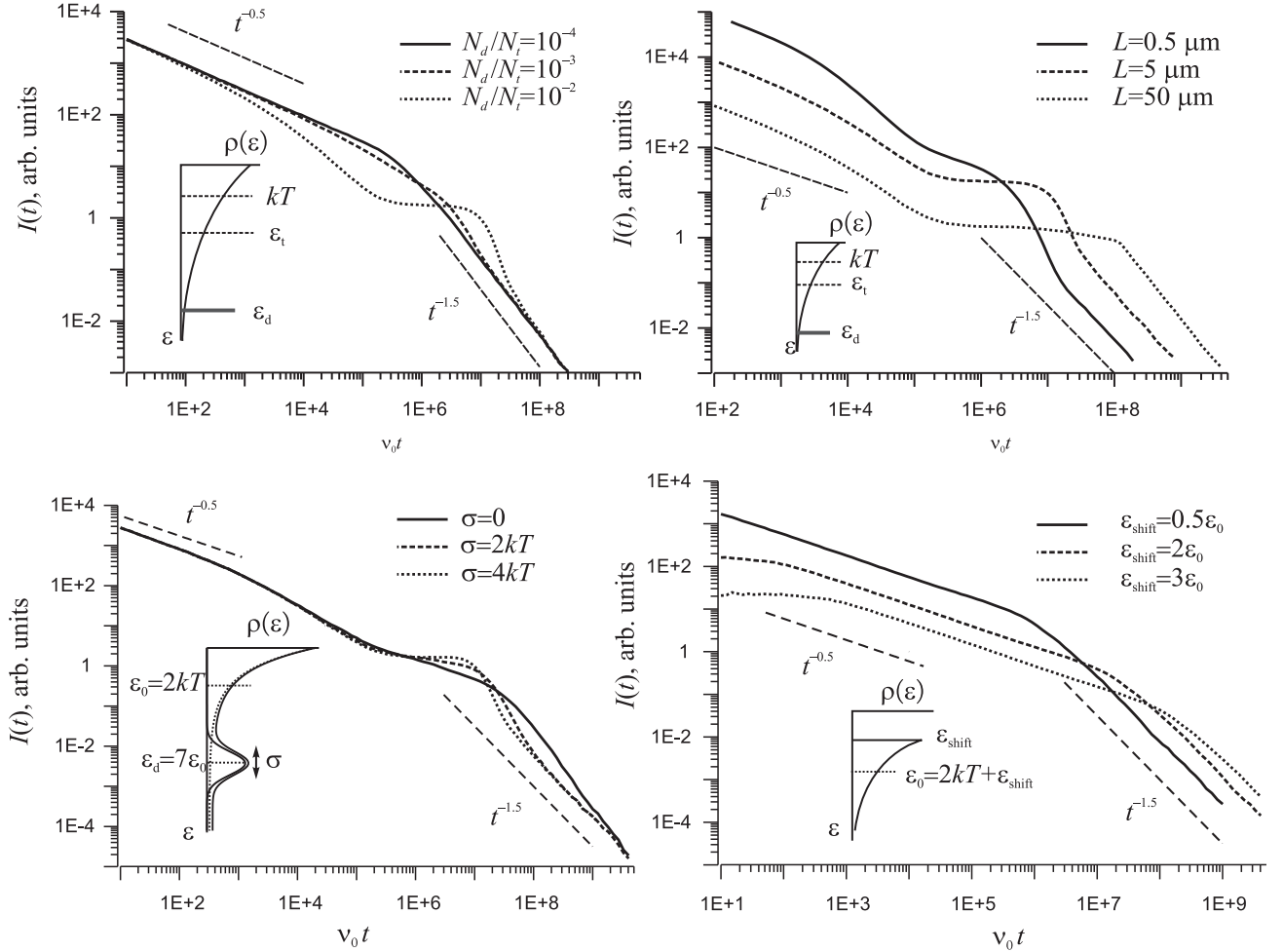}
\caption{Transient current curves in the case of nonmonotonic density of localized states for different situations demonstrated in the insets. $N_t$ and $N_d$ are concentrations of native localized states with exponential density and defects characterized by delta-shaped or Gaussian density, respectively.}\label{fig_nonmono_dos}
\end{figure}

In Fig.~\ref{fig_nonmono_dos}, transient current curves are shown for the case of non-monotonic density of localized states. Details are indicated in the insets. These curves are calculated by the fractional diffusion equation with distributed orders. These calculations confirm the results obtained in~Ref.~\cite{Nik:89}. In particular, the non-monotonic density of localized states leads to the appearance of plateau in the transient current curves. Some new aspects are taken into account: the energetic width of the defect states and the shift below the band edge.

\section{Transient processes in a diode under dispersive transport conditions: Turning on by the current step}

The fact that the fractional differential approach allows us to describe both normal and dispersive transport in terms of the unified formalism can be used for analysis of transients in structures based on disordered semiconductors, by analogy with similar structures based on crystalline semiconductors. We demonstrate this by calculating the transition process in a semiconductor diode under conditions of dispersive transport. In this case, the current $I(t)$ and/or the voltage $U(t)$ play the role of time-dependent transient parameters. The diode performs the transition from the neutral state to the conducting one due to the current step, i.e. the load resistance $R_l$ is substantially greater than the resistance of the diode $R_d$~\cite{Gaman}. On the assumption that low injection conditions are fulfilled, we shall calculate the process for semi-infinite planar diode with $n$-type base. Recombination and generation in the space charge region are neglected. Holes are injected from the $p$-region into the $n$-region with a sharp turn on of current. Later, an equilibrium distribution of holes for a given current step $I_s$ is established as a result of competition between the injection and recombination processes in the base.

The dispersive transport of non-equilibrium holes is described by the generalized diffusion equation
$$
\frac{\partial p_d(\mathbf{r},t)}{\partial t}+ \frac{\tau_l^\alpha}{\tau_d}\ e^{-\gamma_l t}\ _0\textsf{D}_t^\alpha \left[e^{\gamma_l t}\ p_d(\mathbf{r},t)\right]+
\mathrm{div}\left[\mu \mathbf{E}\ p_d(\mathbf{r},t)-D_p\nabla p_d(\mathbf{r},t)\right]+\gamma_d  p_d(\mathbf{r},t)=0.
$$
Here $p_d(\mathbf{r},t)$ is the concentration of non-equilibrium holes.
In the case of one-dimensional diffusion (planar diode), it can be rewritten in the form:
$$
\frac{\partial p_d(x,t)}{\partial t}+ \frac{\tau_l^\alpha}{\tau_d}\ e^{-\gamma_l t}\ _0\textsf{D}_t^\alpha \left[e^{\gamma_l t}\ p_d(x,t)\right]
-D_p\frac{\partial^2 p_d(x,t)}{\partial x^2}+\gamma_d  p_d(x,t)=0.
$$
 Here $\gamma_l$ and $\gamma_d$ are parameters of recombination of localized and quasi-free carriers, respectively. This equation is written for concentration of mobile (quasi-free) carriers, which is applicable in the model of multiple trapping or percolation model
"backbone -- dead ends". Making the Laplace transformation on time yields
$$
 s\tilde{p}_d(x,s)+ \frac{\tau_l^\alpha}{\tau_d}\ (s+\gamma_l)^\alpha \tilde{p}_d(x,s)
-D_p\frac{\partial^2 \tilde{p}_d(x,t)}{\partial x^2}+\gamma_d  \tilde{p}_d(x,s)=p_d(x,0).
$$
Using the evident conditions
$$
p_d(x,0)=0,\quad \lim\limits_{x\rightarrow\infty}p_d(x,t)=0,
$$
and neglecting by time of flight through the spatial charge region of the diode
\begin{equation}\label{eq_diode_tr1}
p_d(0,t)=p_n\left[\exp\left(\frac{e U(t)}{kT}\right)-1\right],
\end{equation}
we obtain solution to this equation in the form
$$
\tilde{p}_d(x,s)=\sqrt{\gamma_d\tau_d+(\gamma_l\tau_l)^\alpha}\ p_n\left[\exp\left(\frac{e U_c}{kT}\right)-1\right]\frac{\exp\left(-x\sqrt{s+\gamma_d+\tau_l^\alpha\tau_d^{-1}(s+\gamma_l)^\alpha }\right)}{s\sqrt{\tau_d s+\gamma_d\tau_d+\tau_l^\alpha(s+\gamma_l)^\alpha }}.
$$
At point $x=0$
$$
\tilde{p}_d(0,s)= p_n\left[\exp\left(\frac{e U_c}{kT}\right)-1\right]\frac{\sqrt{\gamma_d\tau_d+(\gamma_l\tau_l)^\alpha}}{s\sqrt{\tau_d s+\gamma_d\tau_d+\tau_l^\alpha(s+\gamma_l)^\alpha }}.
$$

In the case of dispersive transport, carriers are localized in traps for vast time interval, and one can neglect by recombination of mobile (delocalized) carriers
$$
\gamma_d\tau_d\ll(\gamma_l\tau_l)^\alpha, \quad \tau_d s+\gamma_d\tau_d\ll\tau_l^\alpha(s+\gamma_l)^\alpha.
$$
As a result, we obtain the expression:
$$
\tilde{p}_d(0,s)= p_n\left[\exp\left(\frac{e U_c}{kT}\right)-1\right]\frac{\gamma_l^{\alpha/2}}{s(s+\gamma_l)^{\alpha/2 }}.
$$
Performing the inverse Laplace transformation, we find
$$
p_d(0,t)=p_n\left[\exp\left(\frac{e U_c}{kT}\right)-1\right]\frac{\Gamma(\alpha/2;\gamma_l t)}{\Gamma(\alpha/2)},
$$
where $\Gamma(\nu;t)$ is the incomplete gamma-function \cite{Abram}.
Comparing this relation with Eq.~(\ref{eq_diode_tr1}), we obtain the equation
$$
\left[\exp\left(\frac{e U(t)}{kT}\right)-1\right]= \left[\exp\left(\frac{e U_c}{kT}\right)-1\right]\frac{\Gamma(\alpha/2;\gamma_l t)}{\Gamma(\alpha/2)}.
$$
Solving this equation with respect to $U(t)$ yields
$$
 U(t)=\frac{kT}{e}\ln\left\{1+ \left[\exp\left(\frac{e U_c}{kT}\right)-1\right]\frac{\Gamma(\alpha/2;\gamma_l t)}{\Gamma(\alpha/2)}\right\}.
$$
 \begin{figure}[htb]
 \centering
 \includegraphics[width=0.35\textwidth]{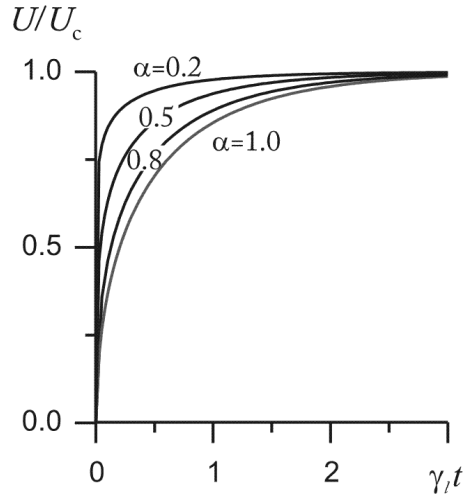}
 \caption{The kinetics of the diode voltage at switching on by the current step under dispersive transport conditions.}\label{fig_tr_process_diode}
 \end{figure}

It is easy to obtain approximate formulas for the two cases
$$
U(t)\approx U_c\  \frac{\Gamma(\alpha/2;\gamma_l t)}{\Gamma(\alpha/2)}, \quad \mbox{for}\quad U_c\ll kT/e,
$$
and
$$
U(t)\approx U_c+\frac{kT}{e}\ln\left(\frac{\Gamma(\alpha/2;\gamma_l t)}{\Gamma(\alpha/2)}\right), \quad \mbox{for}\quad U_c\gg kT/e.
$$

In the case of normal transport $\alpha=1$, and taking into account
$$
\frac{\Gamma(1/2;\gamma_l t)}{\Gamma(1/2)}=\mathrm{erf}(\sqrt{\gamma_l t}),
$$
we arrive at the expression for the diode based on crystalline semiconductors.

Fig.~\ref{fig_tr_process_diode} shows the voltage kinetics for different values of dispersion parameter of holes in the $n$-region, when the diode is switching on by a current step.

\section{On interpretation of the time-of-flight experiment}

Often, the universal form (\ref{anomal}) of the transient current
curves is explained by the exponential density of localized
states~\cite{Mad:88, Tut:05}. For such density of states, multiple trapping and
hopping lead to proportionality $ \alpha \propto T $. In most
experiments, there is observed considerable deviation from this
temperature dependence. Sometimes experimenters do not pay proper attention to this fact and continue to use exponential density of states.

It is known, that the transient curves are very sensitive to the shape of the
energy distribution of traps: the presence of
defect states (even with small concentration) may have a
significant effect. The exponential representation of
localized state density is an evident idealization of more
complicated situations in real disordered semiconductors.
Nevertheless, the transient current curves with two power-type
section are observed more often than one might
expect~\cite{Tut:05}.

Some authors explain the weak dependence of $\alpha$ on $T$ as a
result of topological disorder in these
semiconductors~\cite{Rao02} rather
than the energetic one, as it takes place in cases of multiple
trapping and hopping. Topological disorder can form the
percolation character of the mobility zone and conduction channels. Such phenomena are clearly observed
in porous semiconductors. In this case, the percolation due to the
topological disorder can be described in terms of the fractional
differential kinetics with a temperature-independent dispersion
parameter~\cite{Isichenko}.

Note that the existing analytical approaches to dispersive
transport (Scher-Montroll, Arkhipov-Rudenko, Nikitenko, Tyutnev models) do not
take into account the percolation caused by topological disorder. In
Ref.~\cite{Uch:JETPLett}, we have shown that the fractional version of the diffusion
equation can be derived directly from the universality of transient current curves and the power law dependence of the transient time on the sample
thickness. This means that the equation is valid in
the case of a weak dependence of $ \alpha(T) $. On the other hand,
the comb model of a percolation cluster leads to equations with
fractional derivatives whose orders are temperature-independent in
the case if the correlation length $ \xi $ is
temperature-independent~\cite{Isichenko}.

\begin{figure}[tbh]
\centering
\includegraphics[width=0.7\textwidth]{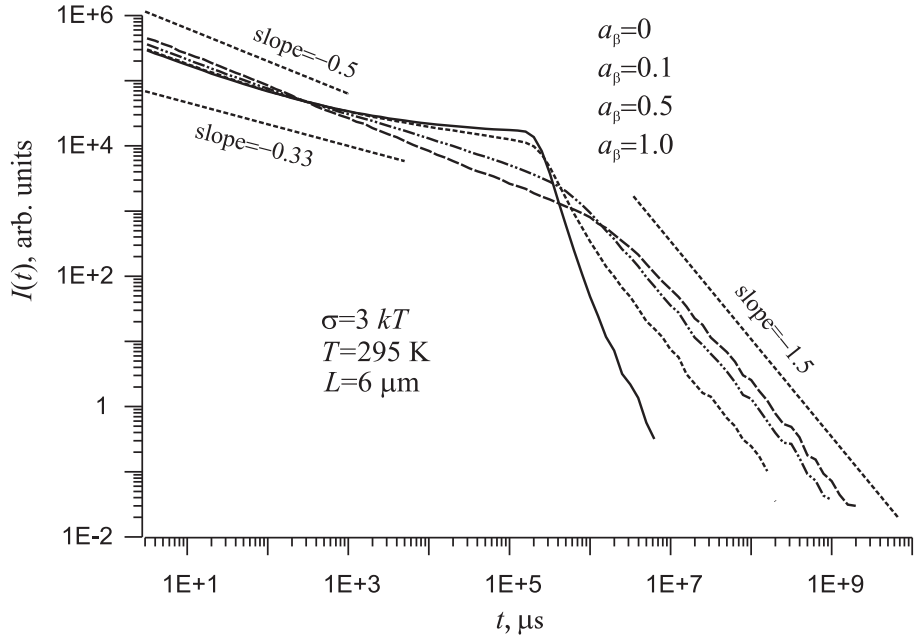}\vspace{-0.3cm}
\caption{Transient current curves in case of Gaussian disorder ($\sigma=3kT$) taking into account percolative nature of conduction channels ($\beta=0.5$), calculated via the solution of equation
(\ref{eq_general2}) for different $a_\beta$ values, $E=10^6$~V/cm, $\tau_0\mu_h E =10$~nm, $\tau_0 D_h=1$~nm$^2$, $c=0.4\cdot10^{6}$~c$^{-1}$.}\label{fig_supress}
 \end{figure}

Since percolation caused by topological disorder and
transfer over the mobility zone are independent processes, we can
write the equation of multiple trapping, taking into account the percolation nature of the
zone~\cite{Uch:12}:
$$
\displaystyle \frac{\gamma}{c_\beta} \ _0\textsf{D}^\beta_t n_d(x,t)+ (1-\gamma)\frac{\partial}{\partial t}
\int_{-\infty}^t n_d(x,\tau)\ Q(t-\tau) d\tau+
$$
\begin{equation}\label{eq_general2}
+\mu E\ \frac{\partial}{\partial x}n_{f}(x,t)-D\frac{\partial^2}{\partial x^2}
n_{f}(x,t)=N\delta(x)\delta(t).
\end{equation}

The first term with the fractional derivative of order $ \beta $
appears due to an asymptotic power law
distribution of the residence time of the carriers in the "dead
branches"\ of percolation cluster. The second term reflects trapping into distributed localized states with arbitrary density.

Fig.~\ref{fig_supress} presents the curves of the transient
current $I(t)$ calculated for the multiple trapping
in states with the Gaussian density $\rho(\varepsilon) \propto \exp (- \varepsilon^2/2 \sigma^2)$.
The current is found by solving equation (\ref{eq_general2})
using the Monte Carlo method, and substituting $ j(x, t) = e \mu E n_d (x, t) $
into the expression for the transient current $I(t)=\int_0^L j(x,t)dx$. The multiple
trapping by traps with the Gaussian DOS without percolation leads to the non-universal curves of the transient current. However, the increase of the
parameter $\gamma$ suppresses the energy disorder and the curves
$I(t)$ takes the universal form.

\begin{figure}[tbh]
\centering
\includegraphics[width=0.7\textwidth]{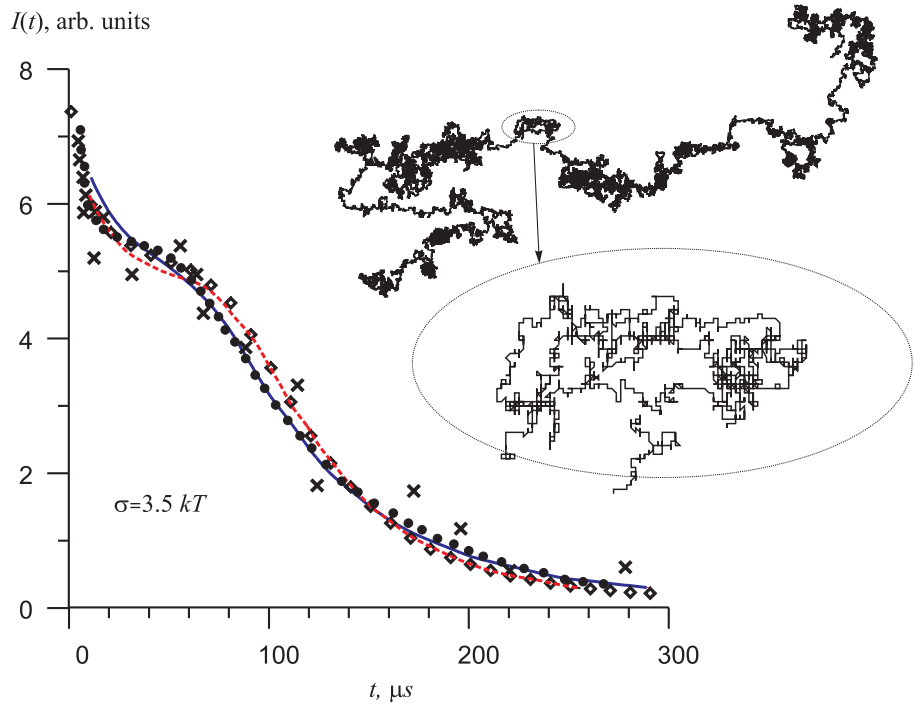}\vspace{-0.3cm}
\caption{Transient current curves: circles are the experimental data for 1,1-bis(di-4-tolilaminophenyl)cyclogexane, digitized from
 [Borsenberger P. M., Pautmeier L., Bassler H., J. Chem. Phys. 94 (1991)], dashed line is the solution from
 [Nikitenko V. R. \& Tyutnev A. P. Semiconductors 41 (2007)], solid line is obtained with the help of equation (\ref{eq_general2}): $\beta=0.5$, $a_\beta=1-a_\alpha=0.1$, $\alpha=kT/\sigma=0.286$, $\gamma=0.98$. The inset demonstrates the trajectory of the 2D-hopping diffusion with parameters $\sigma=3.5 kT=95$~meV, $T$=312~K, $a=0.2d$, $d=1$~nm.}\label{fig_bassler}
\end{figure}

The equation (\ref{eq_FDT_gauss}) similar to Eq.~(\ref{eq_general2}) was obtained for hopping in Ref.~\cite{Sib:12} by involving the transport level concept. Fig.~\ref{fig_bassler} presents the comparison of the calculated transient current curves with experimental
data and results calculated by Nikitenko and Tyutnev\cite{Nik:07} for 1,1-bis
(di-4-tolilaminophenyl) cyclohexane. Noteworthy is the fact that
our simulations of 1D hopping yields the results perfectly consistent
with those obtained from the Nikitenko diffusion equation. This
coincidence is attributed to the fact that  the one-dimensional diffusion equation
with time-dependent coefficients neglects percolation nature of the
trajectories. Involving fractional derivative
of order $\beta$ allows to take into account the nature of
non-Brownian trajectories of hopping particles (see inset in Fig.~\ref{fig_bassler},~a).

\section{Conclusion}

We have presented results obtained in the framework of the fractional differential approach to description of
charge kinetics in disordered semiconductors. The most important property of these processes is their non-Markovity, in other words, the presence of memory. This means
that such kinetics has to be described in terms of integro-differential equations. Self-similarity of these processes leads to fractional kinetic equations~\cite{Kla:85, Barkai3, Barkai, Sib:09, Sta:03, Kla:08, Sok:02, Uch:11}. Some results of this approach concerning transport in disordered semiconductors, samples with spatial distributions of localized states, multilayer structures, transport in non-homogeneous materials with distributed dispersion parameter, and transient process in the diode at dispersive transport conditions are given and discussed in this work.

Concluding, we should stress that
the new approach allows to provide important specifications in interpretation of time-of-flight experiments in disordered semiconductors thanks to the fact that this approach can describe energetic and topological types of disorder in common. Often, shape of transient current curves is explained by a specific density of localized states. For example, the dispersive transport is interpreted in
terms of the exponential density of localized states for inorganic semiconductors (eg $a$-Si:H) and the
Gaussian density for organic semiconductors. In non-organic
semiconductors, multiple trapping is often realized, which is
evidenced by the high mobility of non-equilibrium carriers. As shown
above, the topological disorder can suppress the influence of distributed energy of localized states and lead to "universal"\ curves of
transient current, as in the case of the exponential density of
localized states, but differs from the latter by weak
temperature-dependence of the dispersion parameter. In organic
semiconductors, the charge transfer almost always occurs by hopping and
the percolation nature of the conduction band does not play an essential
role. Thus, the topological form of the mobility edge has to be taken into account in the procedure of reconstruction of the density of localized states from transient current curves. This reconstruction should be performed in close association with the analysis of temperature dependence of current curves. The fractional differential approach forms a mathematical basis for such procedure.

{\small
\section{Acknowledgements}
\noindent
Stimulating discussions with Prof.~S.~Timashev and Prof.~V.~R.~Nikitenko are gratefully
acknowledged. The reported study was partially supported by RFBR (research project~12-01-97031) and the Ministry of Education and Science of the Russian Federation.

{\footnotesize
\section*{Appendix. Table of symbols}

\begin{tabular}{llll}
  $ _0^\alpha\textsf{D}_t $ & fractional Caputo derivative \ & $C$ & anomalous diffusion coefficient \\
  $ _0\textsf{D}^\alpha_t $ & fractional Riemann-Liouville derivative \qquad\qquad \ & $\mathbf{K}$ & anomalous advection coefficient \\
  ${w}_0$ & capture rate into localized states & $\mu$ & mobility of delocalized carriers\\
  $n(\mathbf{r},t)$ & concentration of non-equilibrium electrons &  $n_d(\mathbf{r},t)$ & concentration of delocalized electrons \\
  $p(\mathbf{r},t)$ & concentration of non-equilibrium holes  & $p_d(\mathbf{r},t)$ & concentration of delocalized holes \\
  $D$, $D_p$, $D_n$ & diffusion coefficients for delocalized carriers & $\alpha$ & dispersion parameter \\
  $\tau_0$ & mean delocalization time & $g_+(t;\alpha)$ & one-sided L\'evy stable pdf \\
  $l$ & mean length between localization acts & $\mathbf{E}$ & electric field strength \\
  $\gamma$ & truncation parameter & $\sigma_n$ & conductivity of delocalized electrons \\
  $\gamma_l$ & rate of localized carrier recombination & $\sigma_p$ & conductivity of delocalized holes \\
  $\gamma_d$ & rate of delocalized carrier recombination & $\rho(\varepsilon)$ & density of states \\
  $D_{amb}$ & ambipolar diffusion coefficient   & $\mu_{amb}$ & ambipolar drift mobility \\
  $n_{eff}$ & concentration of electrons at transport level   & $kT$ & Boltzmann temperature  \\
  $I(t)$ & transient current   & $j(x,t)$ & conduction current density  \\
  $L$ & sample thickness & $t_T$ & transient time  \\
  $N$ & number of injected carriers & $\varepsilon$ & energy of localized carriers  \\
  $\psi(t)$ & waiting time pdf &  $U$ & voltage \\
 $\Psi(t)$ & distribution function of waiting times & $\sigma$ & width of Gaussian density of states  \\
 $\overline{\Psi}(t)$ & complementary distribution function  & $s$ & Laplace variable  \\

\end{tabular}
}

}

\end{document}